\documentclass[final,5p,times,twocolumn]{elsarticle}

\usepackage{amssymb}
\usepackage{amsmath}
\usepackage{xcolor}

\journal{Physics Letters B}
\begin{document}
\begin{frontmatter}

\title{Ensuring that toponium is glued, not nailed}

\author{Felipe J. Llanes-Estrada}
\affiliation{organization={Dept. Fisica Teorica and IPARCOS, Univ. Complutense de Madrid},
            addressline={Plaza de las Ciencias 1}, 
            city={Madrid},
            postcode={28040}, 
            state={Madrid},
            country={Spain}}

\begin{abstract}
Hints of toponium might be incipient in LHC data, as given the vast numbers of $t$ quarks produced, some survive on the exponential-decay tail long enough to fasten $t\bar{t}$ together. I here discuss a few differences between the standard Quantum Chromodynamics (QCD) binding (the ``glue'') and exotic short-range binding (the ``nail'').
If the binding energy below threshold reaches the 3 GeV range the peak of the $\eta_t$ is distinct enough that a cross-section dip should be apparent in the line shape, should there only be one isolated resonance, but is filled by the excited QCD states 
adding about a pbarn to the cross section of $t\bar{t}$ production. Their effect for smaller binding energies is a tenuous increase in the cross section.
A new-physics short-range interaction, on the other hand, yields a larger cross-section for equal binding energy (or hardly a visible bound state for similar cross section). This is due to its larger $t\bar{t}$ relative wavefunction at small distances.
Finally, assuming that standard QCD plays out, I comment on what size of constraints on new-physics coefficients one can expect at given precision.
\end{abstract}


\end{frontmatter}

%\tableofcontents


\section{Context: excess cross-section near threshold}
%%%%%%%%%%%%%%%%%%%%%%%%%%%%%%%%%%%%%%%%

The CMS collaboration~\cite{CMS:2024ynj,Jeppe:2024uki} at the LHC has reported an excess cross section of about 7.1(0.8) pbarn in the lowest energy  bin of $m_{t\bar{t}}$ in $t\bar{t}$ production. Neither confirmation nor refutal of this excess by other experiments has yet been presented~\cite{ATLAS:2019hau,ATLAS:2023gsl}.  

It is often said that (with a Bohr time $a_0/(4\alpha_s/3)\simeq 0.33$ fm and a decay time of $\Gamma^{-1}=(2\Gamma_t)^{-1}\simeq 0.067$fm) the $t\bar{t}$ bound state has no time to form. While true for small samples, the large number of $t$ quarks produced at the LHC makes it that a large number of them populate the exponential-decay tail, with substantially longer lifetimes. 
For example, in a recent lepton+jets analysis~\cite{CMS:2023ebf}, the CMS collaboration reports a measurement of the quark mass with 230k top-antitop pairs. Among these, 230k$\int_{0.33}^\infty dt e^{-15t}\simeq 110$ pairs exceed that Bohr-radius time and can thus  be candidates for bound-states. 

\begin{figure}[h]
\includegraphics[width=\columnwidth]{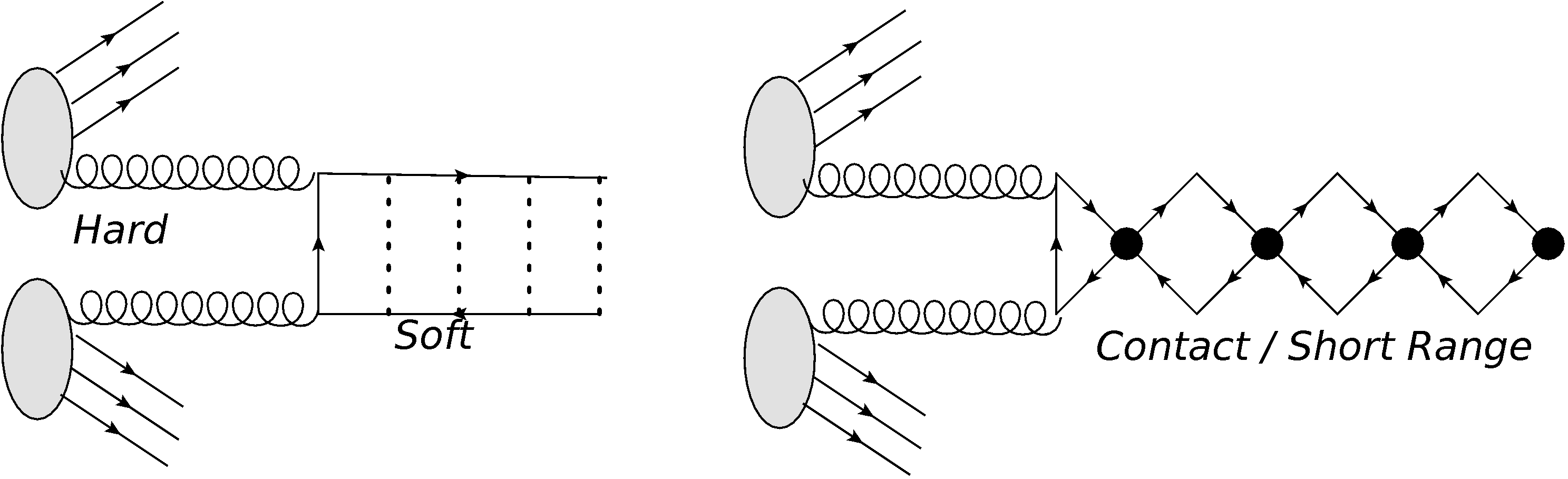}
\caption{\label{fig:Feynman} $\eta_t$ (ground state toponium) production from 
two hard gluons at the LHC. Left: $t\bar{t}$ with binding energy $BE$ from soft gluons with $\mu\sim m_t\alpha_s$, (``glued''). Right: $t\bar{t}$ bound by BSM contact interactions (``nailed''). }
\end{figure}

So before turning to new particles, it is then natural to think about the phenomenology of toponium~\cite{Fuks:2021xje}, and there has been a resurgence of interest in this system~\cite{Aguilar-Saavedra:2024mnm,Yang:2024hnd}. It could also be influenced, in addition to Standard Model forces, by new physics ones, and here I compare the phenomenology of the genuine QCD state bound by gluons and a would-be BSM state bound by contact interactions, as depicted in figure~\ref{fig:Feynman}.

%%%%%%%%%%%%%%%%%%%%%%%%%%%%%%%%%%%%%%%%
\subsection{Sommerfeld effect and/or bound state?}
%%%%%%%%%%%%%%%%%%%%%%%%%%%%%%%%%%%%%%%%
Currently, the only sign of the supposed new state is an enhanced threshold cross section in only one of the major LHC experiments. This could be a sign of a new resonance, but also an enhancement due to other reasons, such as the threshold Sommerfeld enhancement actively invoked in pursuit of new physics~\cite{Petraki:2016cnz,Streuer:2017nsa}. This is caused by an enhanced (for a color singlet) phase-space due to the gluon exchanges marked as ``soft'' in Fig.~\ref{fig:Feynman} (as opposed to the bound states formed below threshold by those exchanges).

\begin{figure}[h!]
\includegraphics[width=0.8\columnwidth]{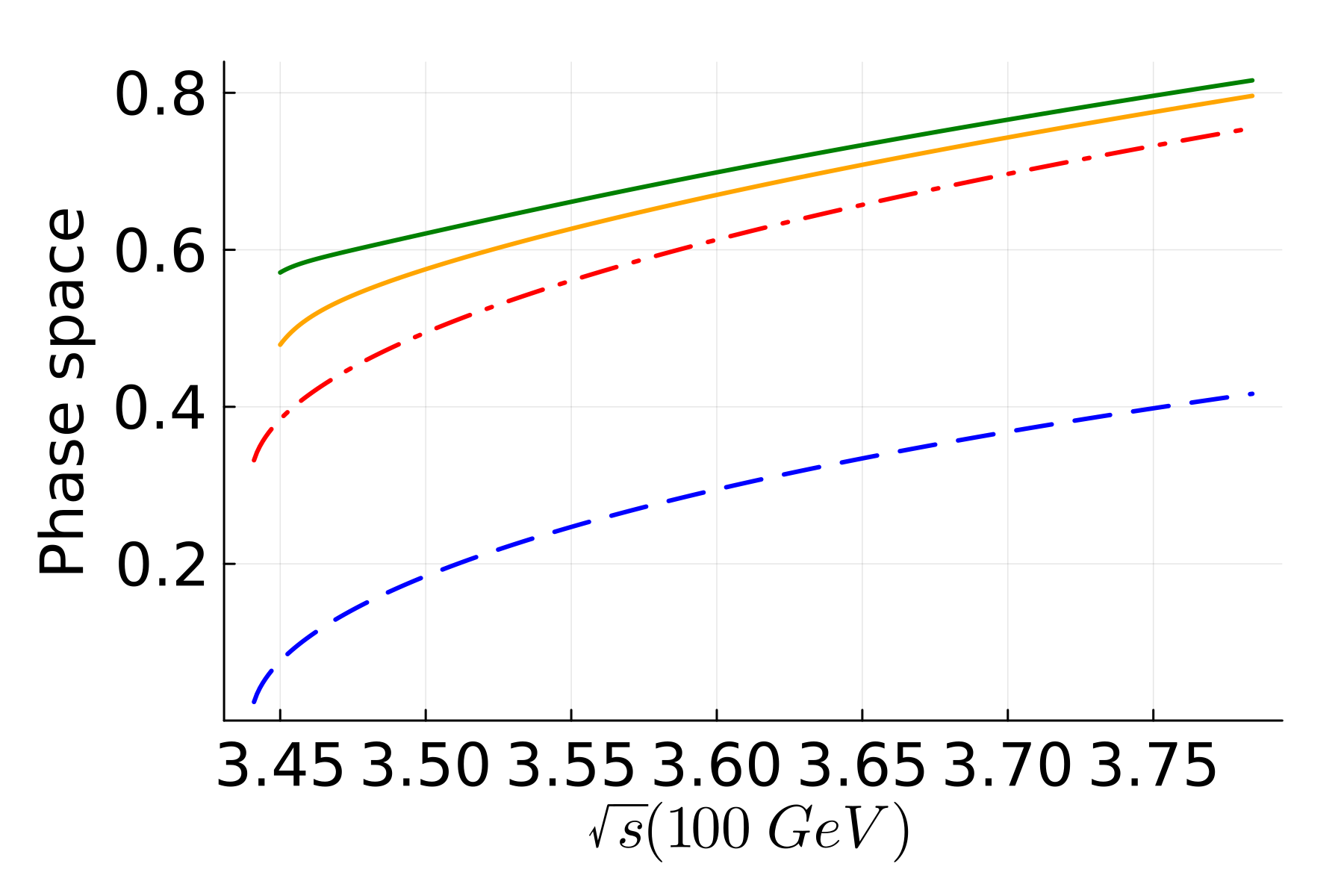}
\caption{\label{fig:Sommerfeld}
Upwards from bottom:
Phase-space factor ($t$-quark velocity $\beta_t$ in a color-singlet $t\bar{t}$ pair);  Sommerfeld-enhanced phase space; effect of including one and then five toponium-like states, all for an ideal narrow $t$ quark, $\Gamma_t=0$. Above 3.5 (100 GeV)$^2$ where $\beta_t\simeq 0.2$ the $t$ quarks become relativistic.}
\end{figure}

To discuss it I first  turn to the study of the production of a $t\bar{t}$ pair near threshold by Fadin, Khoze and Sj\"ostrand~\cite{Fadin:1990wx,Fadin:1991zw,Fadin:1988fn} who computed, extending earlier electrodynamics work by Braun~\cite{Braun:1968njz}, for the colour-singlet pair-production cross section, 
\begin{equation} \label{factorizedXS}
\sigma_{gg\to t\bar{t}}^{s} = \frac{2}{7} \sigma_{gg\to t\bar{t}}^{\rm Born}\times \left[\frac{4\pi}{m_t^2\beta_t} {\rm Im}G_{E+i\Gamma}(0,0)\right]\ .
\end{equation}
The effect of the bracketed term is to substitute 
the phase-space factor
$\beta_t = \sqrt{1-4m_t^2/m_{t\bar{t}}^2} \to \hat{\beta}_t $ as given in Eq.~(\ref{FK})
in the parton-level cross-section near threshold
\begin{equation}\label{production}
\sigma_{gg\to t\bar{t}}^{\rm Born}=\beta_t \frac{7\pi}{192} \frac{\alpha_s^2}{m_{t\bar{t}}^2}\ ,\ \ \ \ \alpha_s(\mu=m_t)\simeq 0.108.
\end{equation}
The enhanced phase-space factor, above threshold for a narrow $t$ quark ($\Gamma_t=0$) is given in Fig.~\ref{fig:Sommerfeld}.  
The lowest (dashed) line represents $\beta_t$ whereas the next (dash-dotted) line includes the Sommerfeld enhancement and is significantly larger.
Recovering the physical width $\Gamma_t$, the integrated cross sections (estimated as described below), $\int dm_{t\bar{t}} \frac{d\sigma}{dm_{t\bar{t}}}$,  are given in table~\ref{tab:cross}. 
$\sigma$ increases from 2.4 pbarn (with the bare phase space $\beta_t$) to about 4.9 pbarn (as computed with $\alpha_s(\alpha_s m_t)\sim 0.147$). This $\sigma$ from the Sommerfeld enhancement being insufficient, the current CMS measurement would already include some additional --perhaps subthreshold-- physics at the 3$\sigma$ level. 
 Therefore, let us turn to the resonant phenomena.

\begin{table}[h!]
\caption{\label{tab:cross} Cross-sections in pbarn around $t\bar{t}$ threshold, for proton-proton collisions at $s_{\rm LHC}=(13 {\rm TeV})^2$, integrating between $m_{t\bar{t}}=$330 and 350 GeV  the differential cross-section of figure~\ref{dipornodip} (from Eq.~(\ref{xsec})). The entries labelled ``short range'' include the new physics potential alone, in this case with a 13 GeV Yukawa mass scale, with no gluon exchange.} 
\centering
\begin{tabular}{|c|c|c|c|}\hline
$\Gamma_t$ & Sommerfeld &  $+\eta_t$ only  & All Bohr \\
only       & enhanced   & ($BE$ 2.5 GeV)   & states  \\ \hline
2.4        & 4.9        & 7.5              & 8.3     \\ \hline
All Bohr   &Short range: & Resonant         & Resonant \\ 
NNLO       & Unresonant & ($BE$ 1 GeV)     & ($BE$ 2 GeV) \\ \hline
9.1        & 7 & 8.5 & 23\\ \hline
\end{tabular}
\end{table}

%%%%%%%%%%%%%%%%%%%%%%%%%%%%%%%%%%%%%%%%
\section{What is the $\eta_t$ binding energy? Scale dependence}
%%%%%%%%%%%%%%%%%%%%%%%%%%%%%%%%%%%%%%%%

The binding energy and total mass are, at leading order,
\begin{equation}
-BE = - \frac{4}{9} m_t \alpha_s(\mu)^2\ \ \ \ \ \ \textrm{or}\ \ \ \ \    M = 2m_t \left( 1 - \frac{2}{9} \alpha_s^2(\mu)
\right)\ .
\end{equation}
This, and any low-order calculation, is affected by the choice of scale. 
If $\mu$ is fixed equal to the ``hard'' one in the parton cross-section of Eq.~(\ref{factorizedXS}),
large logarithms should creep-in as 
$\alpha_s(172{\rm GeV})\times \log((172{\rm GeV}/2{\rm GeV})^2)\simeq 1$ suggesting the need for resummations.

Potential Nonrelativistic QCD~\cite{Brambilla:2004jw} on the other hand indicates that the ``soft'' scale $\mu \sim \alpha_s m_t $ is natural~\cite{Vairo:2021nvu} for the binding gluons, for which, self-consistently, $\alpha_s(m_t\alpha_s\simeq 25 {\rm GeV})\simeq 0.147$.  More virtual gluons are integrated out and are irrelevant at the level of the potential theory.

The resulting binding energy is affected by the scale choice, as shown shortly in table~\ref{tab:spectrum}.
State of the art computations~\cite{Beneke:2008cr,Beneke:2015kwa,Beneke:2024sfa} concur that between LO and NNNLO, 
$BE\in(2,3)$ GeV, computed with a hard scale, increases with the order of perturbation theory. 
(It should be noted that, in the presence of binding, hence strong, interactions near threshold, the $\eta_t$ pole position in $e^-e^+$ and in $gg$ production is expected to be the same due to Watson's final-state interaction theorem and similar results).

Discussing the mass value to GeV precision may seem rather academic at this point in which the experimental binning is at the 20 GeV level, but it should be pointed out that the cross-section excess is quite sensitive to the binding energy, with an excess of about 
7 pbarn if $BE$ is around 2 GeV, but quickly increasing if it reaches 3 GeV. Thus, a good determination of the cross section can be sensitive to the mass of the $\eta_t$.

%%%%%%%%%%%%%%%%%%%%%%%%%%%%%%%%%%%%%%%%
\section{Where are the excited states?}
%%%%%%%%%%%%%%%%%%%%%%%%%%%%%%%%%%%%%%%%
 \begin{figure}[h!]
    \centering
\includegraphics[width=0.8\columnwidth]{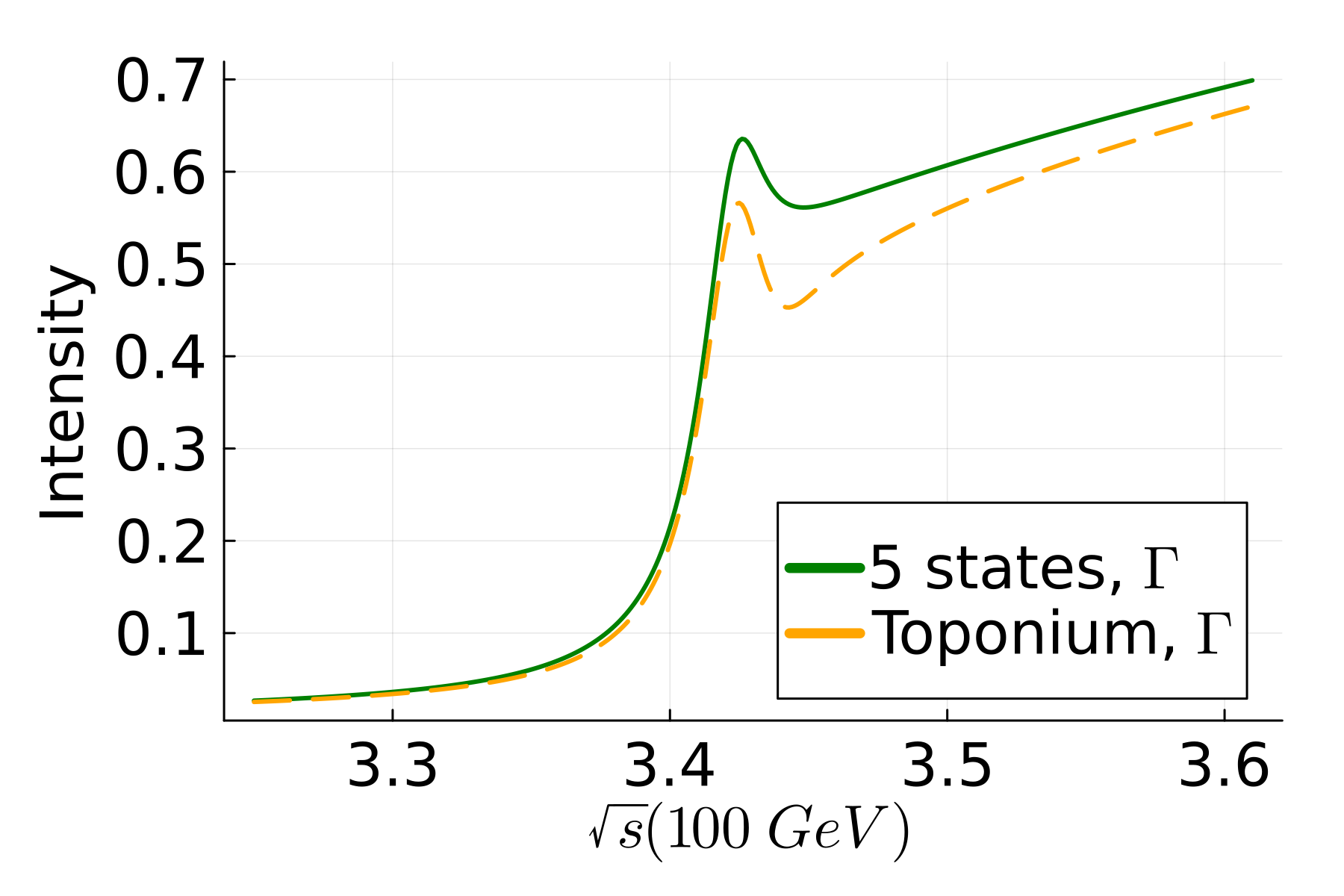}
    \caption{Factor substituting the $\beta_t$ phase-space one for the color-singlet $t\bar{t}$ channel,  including the Sommerfeld enhancement and either one or five toponia states (since they cluster near threshold, this eliminates any dip in the cross-section which appears if the binding energy approaches 3 GeV).}
    \label{fig:phasespacetoponium}
\end{figure}

The Bohr spectrum of a Coulomb potential does include the
excited $s$-wave states $\eta_t^n$ with decreasing binding energy $BE/n^2$ and decreasing coupling to two gluons. 
The QCD interaction is less Coulombic at larger distances due to logarithms and eventually nonperturbative effects as captured by the Richardson or Cornell potentials. A quick recalculation of the spectrum of $\eta_t$ is shown in table~\ref{tab:spectrum} for varying orders of the potential.  (Not shown is the effect of the string tension in the Cornell potential, which eventually pushes excited states above threshold).

\begin{table}[h]
    \centering
        \caption{$|BE|$ spectrum  (in GeV) for a few eigenstates  of the Schr\"odinger equation with the Coulomb potential (for which $BE=0$ is an accumulation point with infinitely many states), and the pNRQCD static potentials at NLO and NNLO. The renormalization scale is soft $\mu\simeq 25$ GeV. Shown as raised/lowered values are the masses corresponding to the scale variation  $\mu/3$ to $3\mu$.  \label{tab:spectrum}}
\begin{tabular}{|c|ccccc|}\hline
LO     & $1.65^{2.64}_{1.14}$&$0.41^{0.66}_{0.28}$&$0.18^{0.29}_{0.13}$&$0.10^{0.17}_{0.071}$&$0.066^{0.11}_{0.031}$\\ \hline
NLO    & $2.26^{2.48}_{1.96}$&$0.76^{0.93}_{0.62}$&$0.40^{0.52}_{0.31}$&$0.25^{0.34}_{0.19}$ &$0.17^{0.24}_{0.13}$  \\ \hline
NNLO   & $2.54^{3.16}_{2.24}$&$0.92^{1.2}_{0.78}$ &$0.52^{0.66}_{0.43}$&$0.35^{0.45}_{0.28}$ &$0.25^{0.34}_{0.20}$   \\ \hline
\end{tabular}
\end{table}
It is clear that in addition to the ground state toponium, at least another four states are around or below the $t\bar{t}$ threshold, close enough to be Coulombic, and I employ them in computing the line shape $\sigma(m_{t\bar{t}})$ from Eq.~(\ref{modphasespace}). This is in agreement with a recent Bethe-Salpeter reassessment~\cite{Wang:2024hzd} where all three $\eta_t$ states reported are below  threshold. 

The effect of including these excited states is most visible in the line shape (Fig.~\ref{fig:phasespacetoponium}) when/if Rayleigh's criterion for separating toponium from threshold, $BE>\Gamma_{t\bar{t}}=2\Gamma_t\simeq 3$ GeV is met. Then a dip between the main $\eta_t$ peak and the continuum would appear, which is filled by the excited states (not individually resolvable). If on the other hand $BE<<\Gamma_{t\bar{t}}$, the only effect of these excitations is a slight increase of the LHC production cross section at the level of at most 1 pbarn  (table~\ref{tab:cross}). 

\begin{figure}[h!]
\includegraphics[width=0.8\columnwidth]{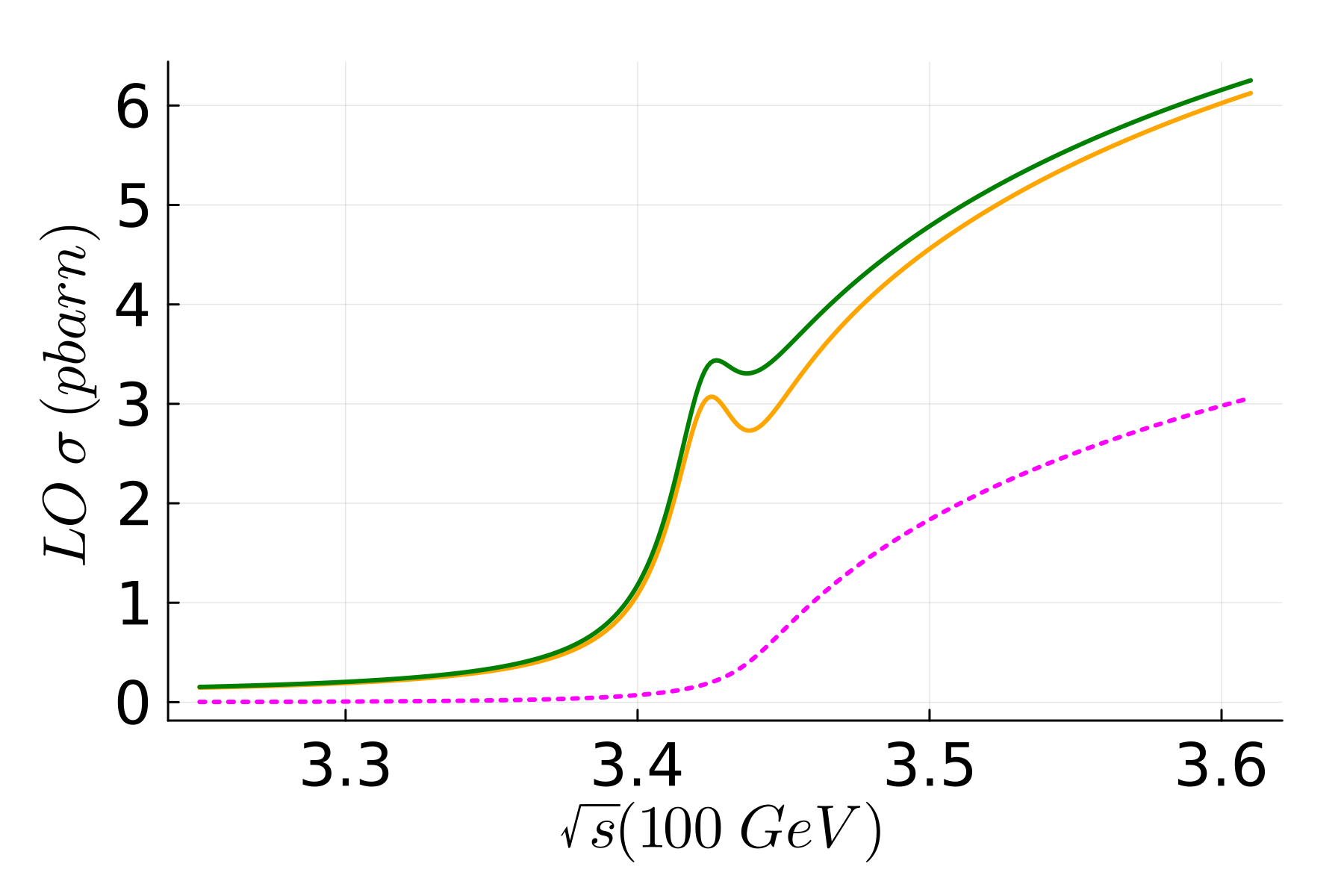}
\caption{\label{fig:partoncross}  Estimate of the parton-level $gg\to t\bar{t}$ cross section 
as function of $m_{t\bar{t}}$ showing the $\eta_t$ peak with binding energy $BE=4$ GeV $>\Gamma_t\simeq 3$ GeV (middle, dashed line). Because of the sufficient separation from threshold, the excited states can be recognized by partially filling the dip (solid line above, includes $\eta_t$ and four excited states near threshold): this is however not completely filled because of the opening of the octet channel, the lowest line. }
\end{figure}

The leading-order parton-level cross section is shown in Fig.~\ref{fig:partoncross} with the $\eta_t$ barely separating from the nominal threshold (here at 344 GeV). 
Interestingly, the decay width $\Gamma_{t\bar{t}}$ does not measurably affect the position of the $\eta_t$, that peaks at around 2.5 GeV of binding energy below the nominal threshold at NNLO with the soft scale choice.

\begin{figure}[h!]
\includegraphics[width=0.8\columnwidth]{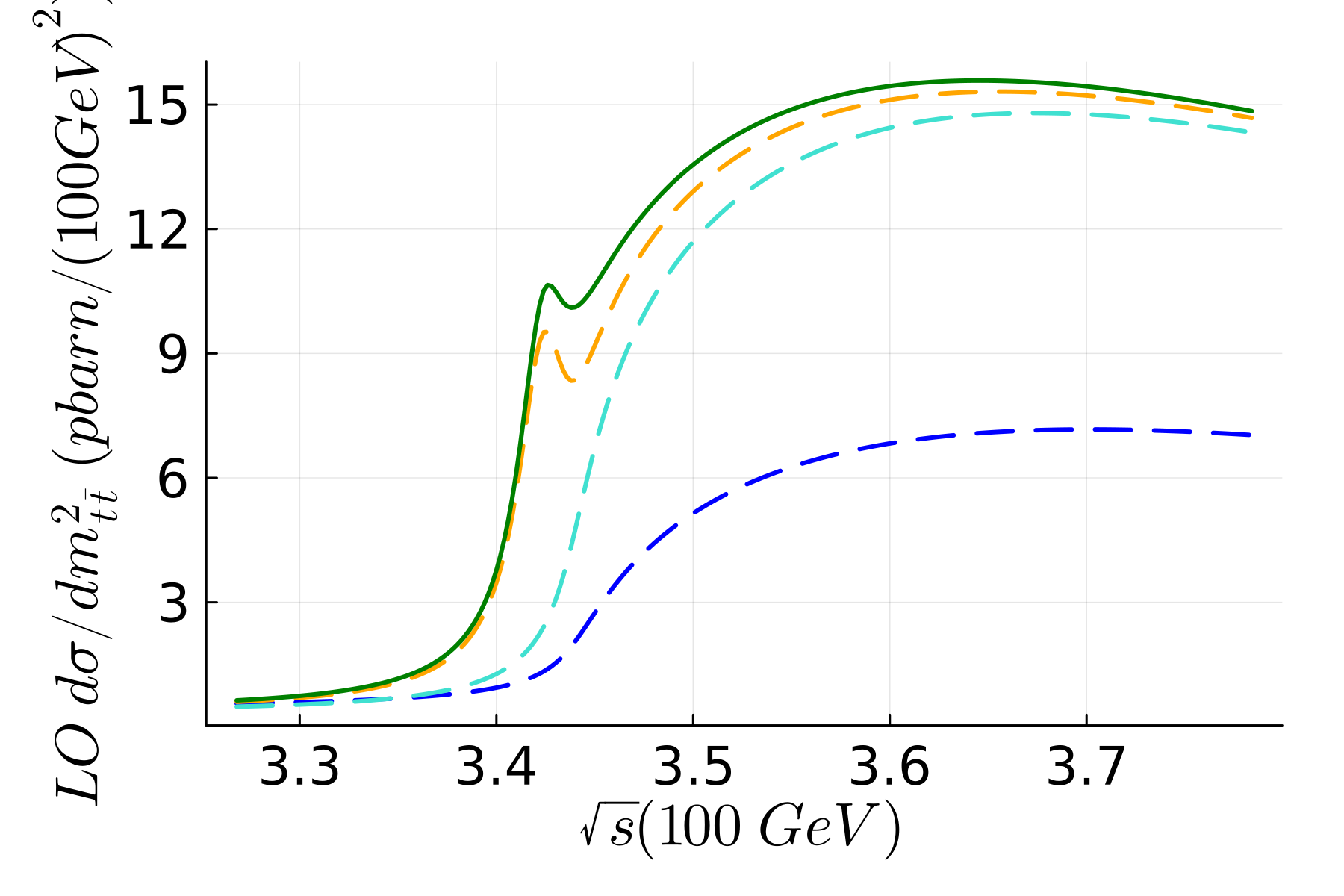}
\caption{\label{dipornodip} Gross estimate of the $pp$ differential cross-sections as functions of $m_{t\bar{t}}^2$. From bottom to top, the would-be production of $t\bar{t}$ with no rescattering; the effect of  the Sommerfeld enhancement; the addition of one $\eta_t$ toponium under the nominal threshold; and the addition of the first $n=1\dots 5$ $\eta_t$ states.}
\end{figure}

The proton-proton cross section is then depicted in figure~\ref{dipornodip}. There, the presence of several toponium excitations practically closes the threshold dip due to the separation of the $\eta_t$ from $m_t+m_{\bar{t}}$, which is still visible in the line shape even in proton-proton collisions, and implies the announced additional pbarn of total cross section as seen in table~\ref{tab:cross}. Should a lepton collider run at the top threshold and obtain the lineshape, or the calorimetric energy resolution at hadron colliders significantly improve, searching for the presence of that dip (isolate state) or its absence (conventional $\eta_t$ tower) could be instructive.

%%%%%%%%%%%%%%%%%%%%%%%%%%%%%%%%%%%%%%%%
\section{Binding by a contact interaction?}
%%%%%%%%%%%%%%%%%%%%%%%%%%%%%%%%%%%%%%%%

Other researchers have studied the possibility of a resonance caused by a new field decaying to $t\bar{t}$. I here concentrate on whether one can exclude a genuine $t\bar{t}$ state bound by a new force alien to QCD. Detailed analysis on new topphilic forces, though not concentrated on toponium, have been reported~\cite{Maltoni:2024wyh}. The extreme case of a new short-range force affecting the $t\bar{t}$ system is a contact operator, which leads to Dirac delta-function potential, 
\begin{equation}
H= -\frac{1}{2m_t/2} \nabla^2 + v \delta^{(3)}(r)\ .
\end{equation}

\begin{figure}[h!]
    \centering
    \includegraphics[width=0.72\columnwidth]{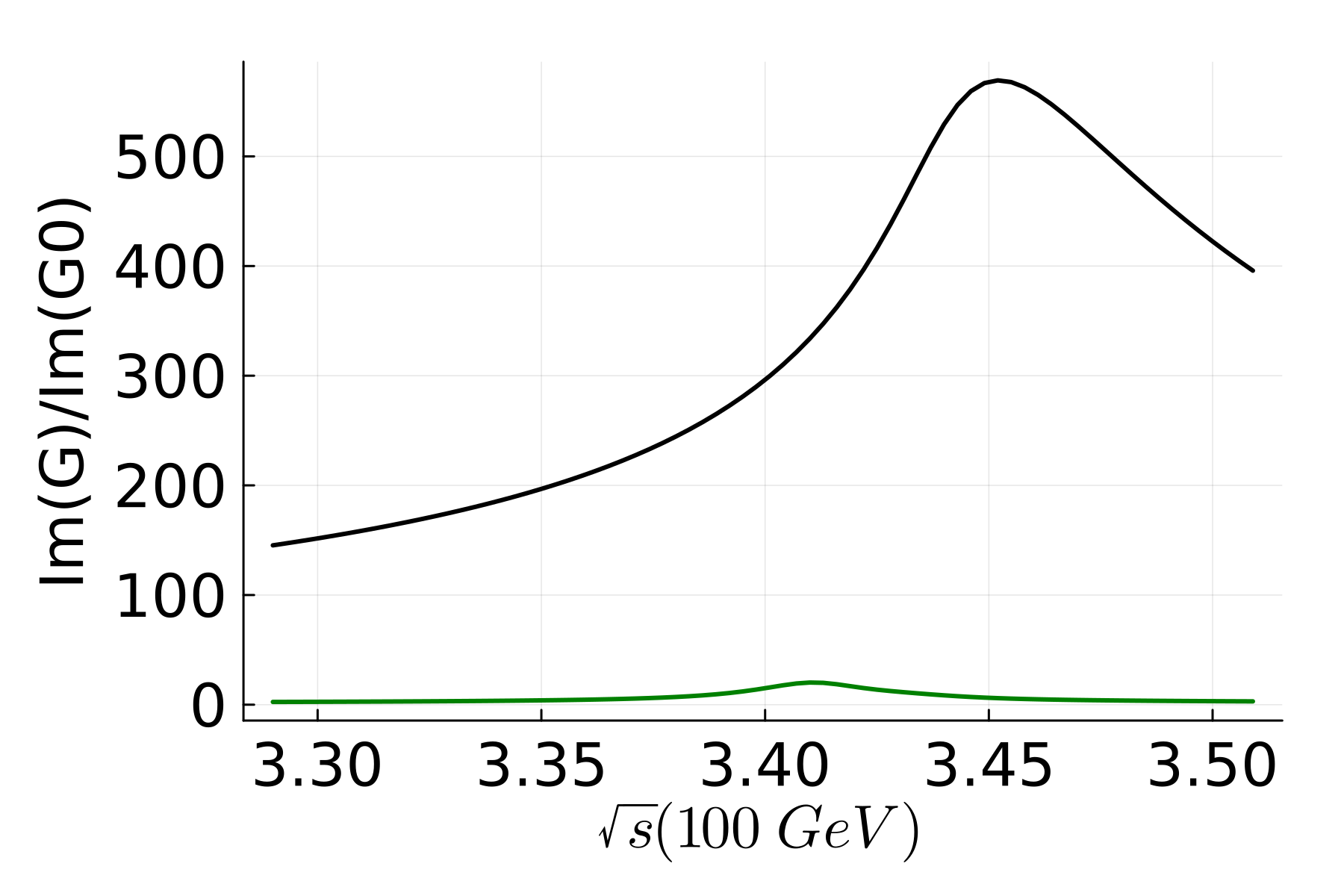}\vspace{-0.2cm}
    \includegraphics[width=0.72\columnwidth]{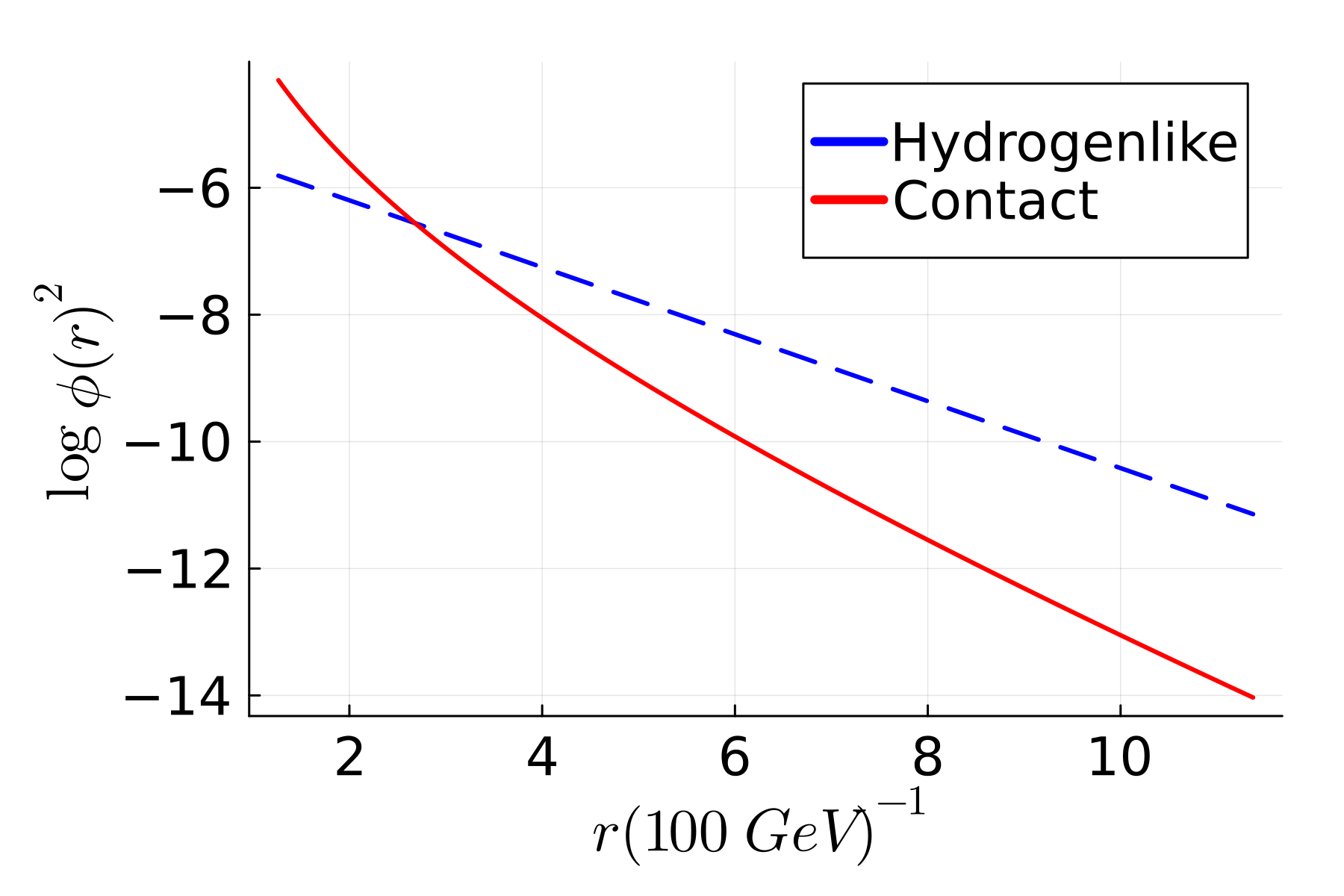}
    \caption{Upper plot: a quasicontact regulated potential with scale $\Lambda=1$ TeV needs such a large coefficient $C(\Lambda)$ to bind a state with energy around 2.5 GeV that the cross section would be hugely enhanced respect to the NNLO computation in QCD (lower line). This is all but ruled out by recent CMS and other measurements. (For a short-range potential with Yukawa mass 13 GeV see Fig.~\ref{fig:shortrange} below). Lower plot: a short range potential has a wavefunction $\phi$ much more concentrated near the origin of $r$ than a Coulomb-like interaction as given by QCD. This causes the larger peak and $\sigma$ at  binding energies similar to the Coulomb one.}
    \label{fig:contact}
\end{figure}

With such singular potentials, the bare Hamiltonian is not bound below~\cite{galindo:1989} and needs renormalization.
This can be achieved~\cite{Jackiw:1991je} by imposing a cutoff $\Lambda$ and absorbing it in the coupling constant as customary,
\begin{equation}
g=v\times \left[1+\frac{m_t\Lambda v}{\pi^2}\right]^{-1}\ .
\end{equation}

The binding energy of the lonely bound state (\cite{Jackiw:1991je} see also~\cite{Gosdzinsky:1990vz})
 that the contact potential supports is then
\begin{equation}\label{boundEcontact}
\textrm{BE} = \frac{1}{m_t^3} \frac{\pi^2}{2g^2}
\end{equation}
which gives a pole in the denominator of the interacting Green's function~\cite{Cavalcanti:1998jx}.
This equation basically determines $g$ from the bound state.
The potential can also be written, extracting the new physics mass scale to make the coupling dimensionless, as
\begin{equation}
V(\mathbf{r}) = \frac{\hat{g}^2}{M^2}\delta^{(3)}(\mathbf{r})\ .
\end{equation}
Here, the Wilson coefficients $C_i$ of the new physics operators~\cite{Alonso-Valero:2024jim} yield the coupling constant $\hat{g}^2$ whereas the effective-theory scale $\Lambda$ is identified with the mass $\leftrightarrow M$ of the new-physics particle, so that $g=C_i/\Lambda^2=\hat{g}^2/M^2$. For numerical work, and also for short- but not zero-range $V$, a regulated form of the contact potential is a Yukawa exchange, 
$V^{\rm BSM} = \frac{\alpha_{\text{contact}}}{|\textbf{r}'-\textbf{r}|}e^{-\Lambda |\textbf{r}'-\textbf{r}|}$ can be used~\cite{Alonso-Valero:2024jim}, where $\alpha_{\rm contact}=C(\Lambda)/(4\pi)$.

\begin{figure}[h]
    \centering
\includegraphics[width=0.7\columnwidth]{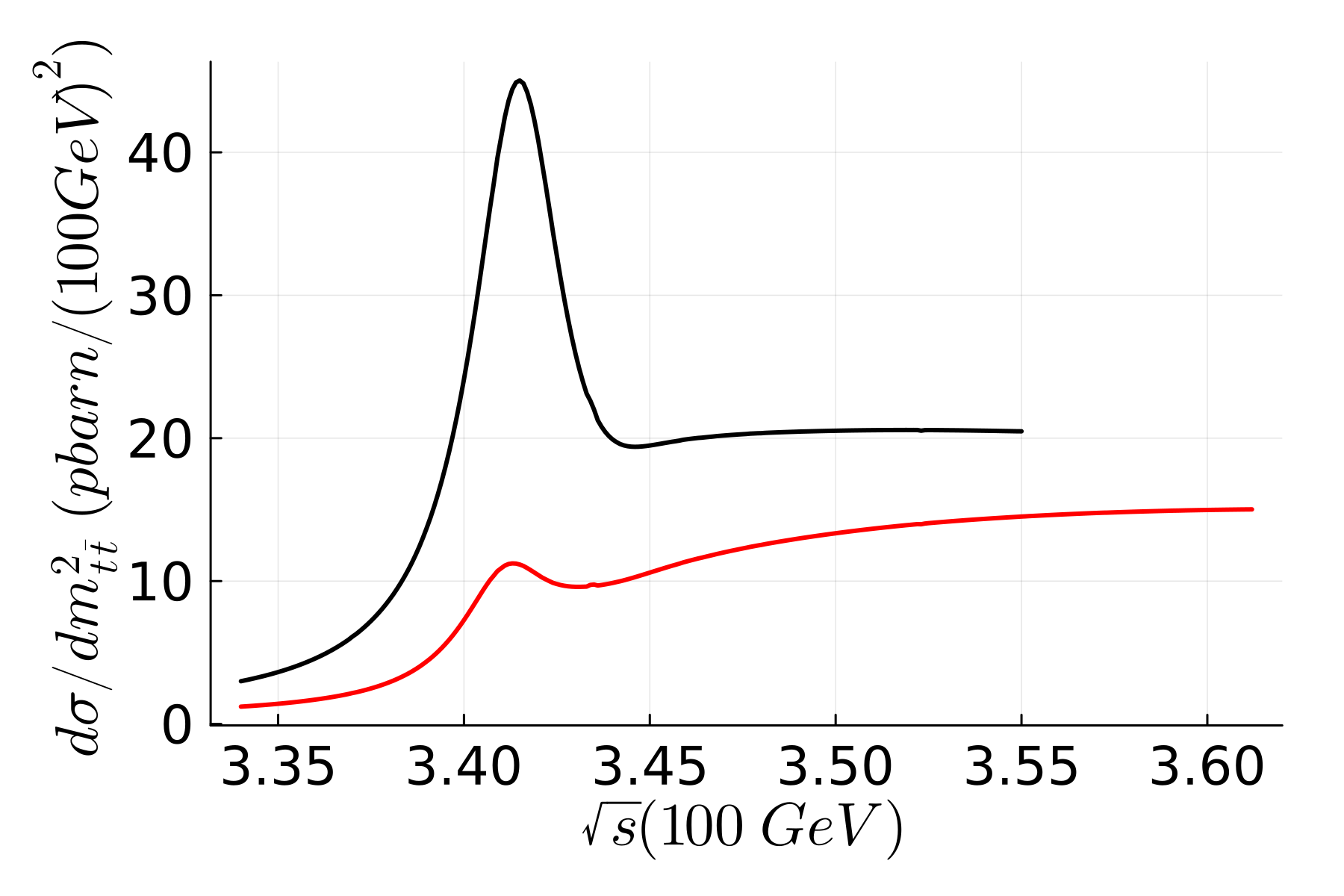}
\includegraphics[width=0.7\columnwidth]{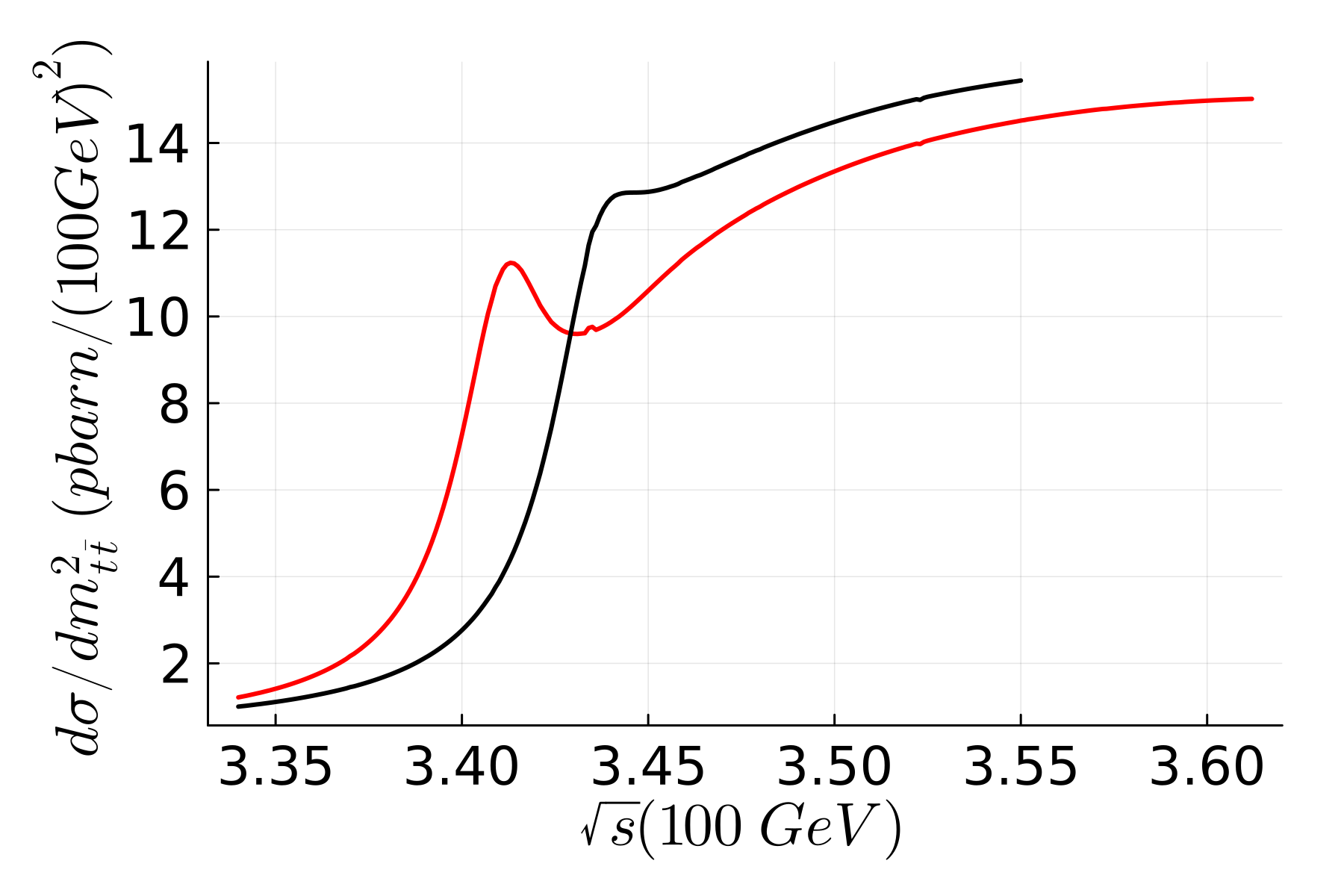}
\includegraphics[width=0.7\columnwidth]{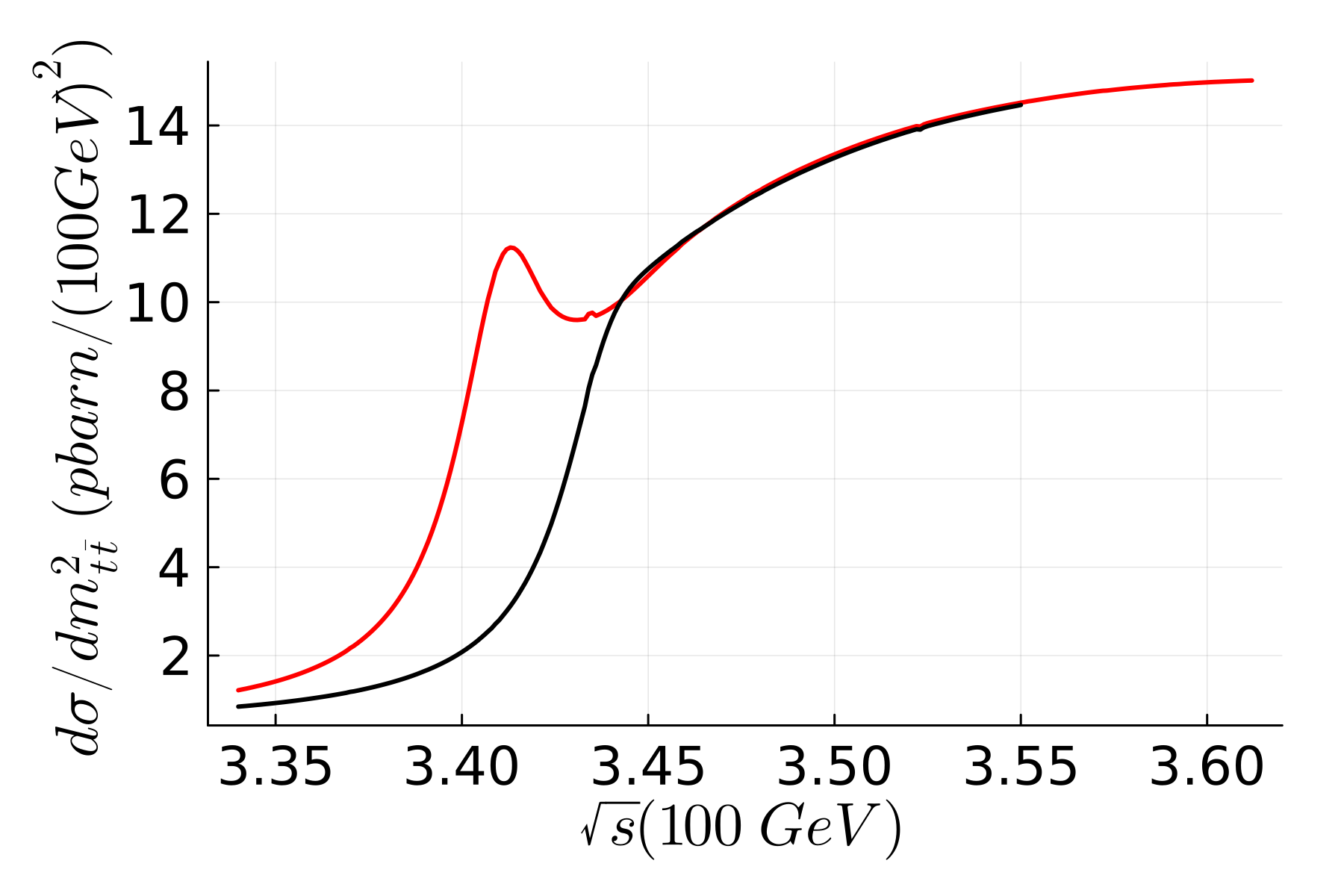}
\caption{\label{fig:shortrange}
The resonant glued-$t\bar{t}$ lineshape (red online) with a soft renormalization scale so that $BE\simeq 2$ GeV, compared with short range potentials.
Upper plot: if the new physics interaction produces one clear bound state very near threshold, akin to the $\eta_t$, in exchange the cross section is much larger. 
Lower plot: the short-range potential is matched to the cross-section above threshold, but then it is too weak for a visible resonance; here $C(\Lambda=M)= 3.7$, $M=13$ GeV, so that $m_t C(\Lambda)/(4\pi) \simeq 47 > 2M=26$ GeV satisfies Bargmann's and other known bounds~\cite{Calogero} guaranteeing a bound state. This however has a meager $BE=1$ GeV and is washed out by the $2\Gamma_t=3$ GeV width.  The middle plot transitions between the other two.
}
\end{figure}

I select for illustration two new physics scales, $M=1$ TeV (genuine contact interactions) and $M=13$ GeV (short range, but resolvable, reported in table~\ref{tab:cross}), with other intermediate cases yielding plausible results omitted. In each case the constant $C(\Lambda=M)$ is tuned to yield a bound state of given energy. Such example cross-sections are shown in figures~\ref{fig:contact} and~\ref{fig:shortrange}.

The outcome of these computations is that, when the contact potential generates bound states with $BE$ similar to QCD's (upper/middle plots in both figures), the cross section is much larger. Conversely, if the cross section is similar, the resonance from a contact potential is lost against the nonresonant production (like in the lower plot of figure~\ref{fig:shortrange}). 
The reason for this behaviour is explained by the lower plot of figure~\ref{fig:contact}: the contact interactions concentrate the bound-state wavefunction much nearer the origin than the Coulomblike ones, and because the relevant Green's function in Eq.~(\ref{factorizedXS}) is evaluated at $r\to 0$, this enhances $\sigma_{\rm contact}/\sigma_{\rm QCD}$. (The Van Royen-Weisskopf ratio among the two wavefunctions $(\phi_{\rm contact}(1/m_t)/\phi_{\rm Coulomb}(1/m_t))^2$ at the hard scale  is of order 40.)

%%%%%%%%%%%%%%%%%%%%%%%%%%%%%%%%%%%%%%%%
\section{How much can we expect to constrain new physics interactions from the bound state?}
%%%%%%%%%%%%%%%%%%%%%%%%%%%%%%%%%%%%%%%%

Let us now adopt the point of view that the CMS excess is indeed caused by the QCD $t\bar{t}$ and see what kind of constraints on new-physics effective theories can be expected.
Recently~\cite{Alonso-Valero:2024jim} we have examined the effect of such contact interactions
(encoding presumed beyond-standard-model (BSM) physics at a higher scale) in multi-$t\bar{t}$ states with the many-body problem treated in Hartree approximation. Because toponium is 
amenable to two-body Schr\"odinger-equation treatment, estimates here are more direct. 
Let us start by the line shape; we can write, 
in perturbation theory, the expected size of the correction to the Coulomb binding energy of a Standard Model toponium given by a new-physics interaction. This is
\begin{equation}
\langle \psi \arrowvert H_{\rm contact} \arrowvert \psi \rangle = 
\int \frac{d^3r}{\pi a^3} e^{-2r/a} g\delta^3(r) = \frac{g}{\pi a^3}\ ,
\end{equation}
and since $a^{-1}=\frac{4\alpha_s}{3} \frac{m_t}{2} \simeq 17$ GeV, with $g\sim C(\Lambda)/\Lambda^2$, 
\begin{equation}
    C(\Lambda) \leq \Delta \langle H_{\rm contact}\rangle_\psi\ / \ (1.56\rm MeV)
\end{equation}
where $\Delta \langle H_{\rm contact}\rangle_\psi$ stands for the precision achievable by a measurement of the $\eta_t$ central mass. If this is of order 1 GeV (50\% of $BE$), the bound on $C$ is of order 600, not competitive with current constraints from global LHC fits. To reach constraints at the level of $O(1-10)$, one needs a  1\% precision on the $\eta_t$ peak position, which does not seem achievable at a hadron collider.

With the only currently available piece of data, the excess threshold $\sigma$, one can still hope to constrain 
the high-energy Wilson coefficients. Adding to the NNLO computation a contact interaction regulated at the TeV scale and with 
Wilson coefficient $K\simeq 10$ it appears one can produce a 1-2 pbarn excess cross-section. This entails that precision below the 10\% level in the experimental excess cross-section may already be enough to gain some sensitivity and will be worth exploring.

%%%%%%%%%%%%%%%%%%%%%%%%%%%%%%%%%%%%%%%%%%%%%%%%%%%
\section{Outlook}
%%%%%%%%%%%%%%%%%%%%%%%%%%%%%%%%%%%%%%%%%%%%%%%%%%%

While I have employed longstanding formulae at leading order in QCD, as well as NLO and NNLO potentials, in the intervening years corrections to the peak position as well as the absolute normalisation of the cross-section have been calculated: for example, in $e^-e^+$ collisions to N$^3$LO; the resonance's maximum here compares very well~\cite{Beneke:2015kwa} to those higher order calculations,  as does the also stable absolute cross-section.  
In the present work, the LO line shape  is surprisingly close to that of~\cite{Beneke:2024sfa} which is a high-order calculation, 
and this calls for a an explanation.  The reason is that the line shape in the $m_{t\bar{t}}$ variable is largely controlled by only three numbers: the $t$ mass marking the threshold, its width $\Gamma_t$, and the $\eta_t$ mass.  

Beneke and Kiyo choose to evaluate $\alpha_s(\mu)$ at a hard scale of order 100 GeV, so that $\alpha_s\sim 0.12-0.13$, and the high-order calculation achieves a good measure of scale-independence. At lower orders in perturbation theory, it is important that the scales are well set, since the result is renormalization-scale dependent. The judicious choices of $\mu=O(m_t)$ for the hard parton level production and $\mu=O(\alpha_s m_t)$, for the bound state, capture much of the physics~\cite{Brodsky:1982gc,DiGiustino:2023jiq} without resort to extremely technical field-theory computations (at the expense of having to set the two scales separately, which may be aesthetically unpleasing, but practical).

I have shown that \emph{(i)} the characteristic excited $\eta_t$ spectrum due to glued toponium manifests itself as closing the gap between the main $\eta_t$ peak and the continuum even if the $\eta_t$ is separated from the threshold (remember that a contact interaction, on the contrary, has only one bound state); \emph{(ii)} that if the $\eta_t$ is near threshold, with no gap to fill, the only manifestation of the excited states is a small enhancement of $\sigma_{pp\to t\bar{t}}$; \emph{(iii)} the cross-section excess reported by CMS is compatible with SM $t\bar{t}$ toponium glued by the QCD force, and if a new short-range interaction is invoked to explain it or a fraction thereof, it will not display a bound state as the excess is too small; \emph{(iv)} conversely, if an $\eta_t$ like state is isolated, the reported cross section is too small for a state bound by contact interactions (which would need large Wilson coefficients, also in severe tension with Higgs Effective Field Theory bounds~\cite{Alonso-Valero:2024jim}); and \emph{(v)} extraordinary precision would probably be needed in the mass determination to improve constraints on contact effective theories.

%%%%%%%%%%%%%%%%%%%%%%%%%%%%%%%%%%%%%%%%%%%%%%%%%%%
\section*{Acknowledgements}
%%%%%%%%%%%%%%%%%%%%%%%%%%%%%%%%%%%%%%%%%%%%%%%%%%%
This report was informed by the November 2024 meeting of the LHC-top cross section working group. The author acknowledges comments by F. Maltoni, E. Palencia, S. Tentori, and other participants, as well as the hospitality of the CERN-TH members. 
Supported by grants PRX23/00225 (estancias en el extranjero) and PID2022-137003NB-
I00 of the Spanish MCIN/AEI /10.13039/501100011033/

%%%%%%%%%%%%%%%%%%%%%%%%%%%%%%%%%%%%%%%%%%%%%%%%%%%
\appendix
\section{Calculational detail}
%%%%%%%%%%%%%%%%%%%%%%%%%%%%%%%%%%%%%%%%%%%%%%%%%%%

\paragraph{Proton-proton cross sections}

To obtain ballpark estimates of the relevant proton-proton cross-sections we need to fold the parton-level ones with the gluon luminosity $ \frac{d\mathcal{L}}{dm^2_{t\bar{t}}}$, 
\begin{eqnarray} \label{xsec}
\frac{d\sigma}{dm^2_{t\bar{t}}} 
% &=& \hat{\sigma}_{gg\to tt}  \nonumber \\  &=& 
=\hat{\sigma}_{gg\to tt} \int_{m^2_{t\bar{t}}}^1
\frac{dx}{x s_{\rm LHC}} f_g(x) f_g\left( \frac{m^2_{t\bar{t}}}{x s_{\rm LHC}}\right) \ .
\end{eqnarray}
That gluon luminosity  is based~\cite{Karmour,Lai:1999wy} on the simple parametrization family $\frac{1}{N}x^a(1-x)^b$ with numerical values slightly different at low and high-$x$; 
\begin{equation} \label{partons}
    f_g(x) = \left\{
\begin{tabular}{cc}
$\frac{1}{4.28} x^{-1.77} (1-x)^{6.06} $ & if $x< 0.1$ \\
$\frac{1}{4.71}x^{-1.79}(1-x)^{5.81}$ & if $x\geq 0.1$
\end{tabular}
    \right.\ .
\end{equation}
(Full Monte Carlo simulation of $t\bar{t}$ production~\cite{Fuks:2021xje} is beyond my scope here.)

\paragraph{Static potentials}
 The $t$ quark mass adopted is $m_t=172$ GeV~\cite{Garzelli:2023rvx}, 
and my considerations are little sensitive to this choice as the important number here is $BE=m_{\eta_t}-2m_t$ which we hope to experimentally know at $O(1)$ uncertainty for now. When/if the precision in that binding energy reaches the few percent level, one can then carefully study the scheme dependence. A nonrelativistic treatment near threshold is warranted.
The potentials to reproduce the spectrum of table~\ref{tab:spectrum} stem from a standard expansion in $\alpha_s$,
\begin{eqnarray}
    V_{\rm NNLO} = -\frac{4}{3} \frac{\alpha_s(\mu)}{r}\left( 1+ f_{\rm NLO}+f_{\rm NNLO}\right)\ .
\end{eqnarray}
For example, the NLO pNRQCD static potential is
\begin{equation} \label{nlo}
    V_{\rm NLO} (r) = \frac{\alpha_s(r^{-1})}{4\pi}\left( a_1 + 2\beta_0 (\gamma_E+\log(\mu r))\right)
\end{equation}
where $\gamma_E \approx 0.57721$ and 
\begin{equation}
    a_1 = \frac{31}{3} - \frac{10N_f}{9}\ ,\ \  \ \
    \beta_0 = 11 - \frac{2N_f}{3}
\end{equation}
$N_f=5$ being the number of flavors below the soft scale of order 20 GeV.
The NNLO one can be found in~\cite{Llanes-Estrada:2011gwu} and refs. therein.
The running coupling constant is computed from the approximate solution in Eq.~(9.5) of~\cite{ParticleDataGroup:2016lqr} (see also~\cite{Chetyrkin:1997un}) to the renormalization group equation, which generalizes the well known three-flavour, one-loop result~\cite{Prosperi:2006hx}
$ \alpha_s(\mu^2) = \frac{4\pi}{9L}$, $L\equiv \log \frac{\mu^2}{\Lambda_{\rm QCD}^2} $,
and which I deploy through order $1/(\beta_0 L)^3$, 
\begin{eqnarray}\hspace{-0.5cm}
\frac{\alpha_s}{\pi}=\frac{1}{b_0 L}\left( 1-\frac{b_1^2}{b_0}\frac{\log(L)}{L} + \frac{b_1(\log(L)^2-\log(L)-1)+b_0b_2}{b_0^4 L^2}
\right)\hspace{0.1cm}
\end{eqnarray}
with scale $\Lambda_{\rm QCD}=0.22$ GeV. If at any point in the calculation of the spectrum the scale evaluation is too low (due to the vagaries of the diagonalization grid) and $\alpha_s\geq 0.5$, it is immediately saturated to that value 0.5. This should not affect the energy of the lowest states in the pNRQCD understanding and avoids computer glitches.  
The coupling at the hard scale for the production formula Eq.~(\ref{production}) is here $\alpha_s=0.108$.

\paragraph{Average momentum}
In the standard Coulomb wavefunction, $\sqrt{\langle p^2\rangle_{\rm Coulomb}}=\sqrt{m_t BE}$. In the nailed-toponium wavefunction from a contact interaction, $\langle p^2\rangle$ is divergent, and after imposing the regulator $\Lambda$ one can obtain, by evaluating $\langle T\rangle = E-\langle V_{\rm contact}\rangle$, 
$
  \sqrt{\langle p^2\rangle_{\rm contact}}=  \sqrt[4]{m_t BE} \sqrt{\sqrt{m_t BE}+\Lambda \frac{\pi}{2}}\ . 
$

This means that, whereas the Bohr momentum is of order 20 GeV for ordinary ``glued'' toponium, it is of order 200 GeV for ``nailed'' toponium. The Coulomb wavefunction $e^{-r\sqrt{m_t BE}}\sqrt[4]{(m_t  BE)^3}/\sqrt{\pi}$ is then substituted by
\begin{equation}
\phi(r)_{\rm contact} = \frac{\sqrt[4]{m_tBE}}{\sqrt{2\pi}} \frac{e^{-r\sqrt{m_tBE}}}{r} 
\end{equation}
(integrals thereof are to be clipped by the  regulating scale $\Lambda$.)

\paragraph{Coulomblike modified phase space}
Following Fadin and Khoze~\cite{Fadin:1991zw}, the $t$-quark velocity which provides the production phase space 
picks up a Sommerfeld Coulomb threshold enhancement 
(first two terms) and a sum over bound states,
\begin{eqnarray} \label{modphasespace}
\hat{\beta}_t=\sqrt{1-\frac{4m_t^2}{m_{t\bar{t}}^2}}\longrightarrow \frac{p_+}{m_t}+\frac{2p_0}{m_t} {\rm atan} \left(\frac{p_+}{p_-} \right)+ \nonumber\\ \label{FK}
+\frac{2p_0^2}{m_t^2}\sum_{n=1}^{n_{\rm max}}\frac{1}{n^4}   \frac{\Gamma_t p_0 n + p_+(n^2 \sqrt{E^2+\Gamma_t^2} +p_0^2/m_t )}{(E+p_0^2/(m_t n^2))^2+\Gamma_t^2} 
\end{eqnarray}
The momenta therein are Bohr's $p_0\equiv \frac{4}{3}\alpha_s \frac{m_t}{2} $ and the two  that split from the one characteristic to the $t$-quark width, which are $p_\pm \equiv \sqrt{\frac{m_t}{2}\left( \sqrt{E^2+\Gamma_t^2} \pm E   \right)}$.
The sum is taken up to $n_{\rm max}=5$ which is the number of states near threshold.
To produce figure~\ref{fig:Sommerfeld}, the width $\Gamma_t$ has been set to 0.

The simple substitution of the phase space from
Eq.~(\ref{modphasespace}), that leads to the soft physics providing a multiplicative factor to the parton-level cross-section, is a manifestation of factorization which remains the subject of active investigation~\cite{Brambilla:2024iqg}.

In the absence of interactions nor bound states, excepting the $\Gamma_t$ width, I employ as a reference the Breit-Wigner (BW) spreading of the conventional phase-space, namely
\begin{equation}
\hat{\beta}_t(s) = \int_0^{\sqrt{s}}d\mu_A
\int_0^{\sqrt{s}-\mu_A}d\mu_B \sqrt{1-\frac{(\mu_A+\mu_B)^2}{s}} {\rm BW}(\mu_A^2)
{\rm BW}(\mu_B^2)
\end{equation}
with the relativistic function, in terms of $\gamma\equiv m_t\sqrt{m_t^2+\Gamma_t^2}$,
\begin{equation}
{\rm BW}(s)= \frac{\left(
\frac{\sqrt{8}}{\pi} m_t\Gamma_t \gamma/ \left(\sqrt{m_t^2+\gamma}\right)  \right)}{(s-m_t^2)^2+m_t^2\Gamma_t^2} \ . 
\end{equation}
This is needed, for example, to produce the octet cross-section in figure~\ref{fig:partoncross}, which afterwards can be suppressed by the corresponding Sommerfeld factor.

\paragraph{Modified phase space with arbitrary (particularly, short range) potential}

For generic potentials, trying to obtain explicitly analytical expressions for the needed Green function is taxing, and resort to numerical algorithms~\cite{Strassler:1990nw} common. 
To solve
\begin{equation}
  \left[  \frac{-1}{m_t} \left( \partial_r^2 +\frac{2\partial_r}{r}\right) + V(r-r')-(E+i\Gamma_t) \right] G(r,r')=\frac{\delta(r-r')}{4\pi r^2}
\end{equation}
I discretize the system on a finite grid, also used to diagonalize the Hamiltonian to check the spectrum, with $N_{\rm grid}=100\dots 4000$ and varying long-distance cutoff (a reasonable one for $\eta_t$ is $50/(m_t\alpha_s)$) to check for sensitivities, obtaining a linear system for $G$,
$\sum_{j=1}^{N_{\rm grid}}A_{ij}G_{jk}=D_{ik}$ which we need to solve for $r\to 0$ (i=1) with $D_{1k}$ concentrated in the first $k=1$ element, so the delta function (or a Yukawa if narrow enough) is regulated as a top hat.
Numerical work employs Julia~\cite{Bezanson:2014pyv}.

%%%%%%%%%%%%%%%%%%%%%%%%%%%%%%%%%%%%%%%%%%%%%%%%%%%%%%%%%%

\end{document}